\documentclass[prl]{revtex4}

\usepackage{graphicx}
\usepackage{dcolumn}
\usepackage{bm}
\usepackage{psfig}
\usepackage{CJK}
\usepackage{amsmath}
\usepackage{amssymb}
\usepackage{amsthm}
\usepackage{amsfonts}                      %
\usepackage[english]{babel}                
\usepackage{epstopdf}
\usepackage{color}

\begin{document}
\draft
\title{A universal expression of near-filed/far-field boundary in stratified structures}
\author{Chao Li$^1$, Tengwei Zhang$^1$, Huaiyu Wang$^2$, and Xuehua Wang$^{1,*}$}
\draft

\address{$^1$ State Key Laboratory of Optoelectronic Materials and Technologies, School of Physics and Engineering, Sun Yat-sen University, Guangzhou 510275, China}
\address{$^2$ Department of Physics, Tsinghua University, Beijing 100084, China}

\begin{abstract}

The division of the near-field and far-field zones for
electromagnetic waves is important for simplifying theoretical
calculations and applying far-field results. In this paper, we have
studied the far-field asymptotic behaviors of dipole radiations in
stratified backgrounds and obtained a universal empirical expression
of near-field/far-field (NFFF) boundary. The boundary is mainly
affected by lateral waves, which corresponds to branch point
contributions in Sommerfeld integrals. In a semispace with a higher
refractive index, the NFFF boundary is determined by a dimensional
parameter and usually larger than the operating wavelength by at
least two orders of magnitude. In a semispace with the lowest
refractive index in the structure (usually air), the NFFF boundary
is about ten wavelengths. Moreover, different treatments in the
asymptotic method are discussed and numerically compared. An
equivalence between the field expressions obtained from the
asymptotic method and those from reciprocal theorem is demonstrated.
Our determination of the NFFF boundary will be useful in the fields
such as antenna design, remote sensing, and underwater
communication.

\end{abstract}

\maketitle


\section {1. Introduction}

In many cases, an electromagnetic field shows near-field or
far-field behavior when it is observed in a region near or far from
the field source. The division of the near field and far field is
not only helpful for simplifying theoretical calculation in
different regions, but also easy to highlight the features of the
field in respect regions so as to provide physical interpretations
of the field behaviors. An important question of how to distinguish
the near and far fields naturally arose. Usually, the field can be
expanded by negative powers of the distance between the field point
and source \cite{r01,r02,r03}. The far field is dominated by the
lowest order term, and the near field by the higher-order terms. The
former arises from the requirement of energy conservation and is
call a ``far field approximation'' or ``leading-order approximation
(LOA)''.

For dipole radiations in a vacuum background, the
near-filed/far-field (NFFF) boundary is of the order of light
wavelength $\lambda_{0}$, $L\sim\lambda_{0}$. As the source becomes
larger, this dimension should be corrected as $L\sim
D^{2}/\lambda_{0}$ by the diffraction theory, where $D$ represents
slit widths or antenna lengths. However, what we frequently
encounter in practical situations are stratified backgrounds, rather
than the vacuum one. Stratified backgrounds are of complex
configurations, in which the NFFF boundary has not been properly
addressed yet. To reveal the NFFF boundary in such backgrounds is
not only a complement in theory, but also very important for many
practical applications, such as remote sensing, antenna design, NFFF
transformation and so on \cite{r01,r02,r03,r04,r05,r06,r07,r08,r09}.
In these applications, the distances between sources and observation
points are usually much larger than operating wavelengths, so that
the sources can be treated as dipoles. Here we investigate the
far-field asymptotic behaviors of dipole radiations in stratified
structures, and address a universal NFFF boundary.

Calculation of the dipole radiations in stratified structures is
mathematically equivalent to dealing with Sommerfeld integrals
(SIs). In general, numerical evaluations of the SIs are not easy
since the integrals have an oscillatory feature and possess
singularities along or near the integration paths. Several numerical
methods have been proposed to calculate the SIs so far. Paulus
\emph{et al}. \cite{r10} developed a direct numerical integration
method by appropriately choosing the integration path that avoided
all possible singularities in the complex plane. This method has
been considered as a standard test for the accuracies of other
methods. The discrete complex image method \cite{r11,r12,r13,r14}
and the rational function fitting method \cite{r15,r16} tried to
expand the integrands by simple functions to obtain closed-form
solutions. On the other hand, under the far-field approximation,
closed-form expressions could be obtained from SIs by asymptotic
methods
\cite{r02,r03,r04,r05,r06,r17,r18,r19,r20,r21,r22,r23,r24,r25,r26,r27,r28,
r29,r30,r31} or reciprocal theorem \cite{r32,r33,r34,r35}. Compared
with the numerical ones, the asymptotic methods are of advantages of
simplicity, easy programming, and clear physical interpretations.

The stationary phase method and the steepest descent method are two
commonly used asymptotic methods
\cite{r02,r03,r04,r05,r06,r17,r18,r19,r20,r21,r22,r23,r24,
r25,r26,r27,r28,r29,r30,r31}. Both of them approximate the value of
a rapid-oscillating integral by the contributions around stationary
points (SPs) or saddle points, and provide the same LOA results for
the SIs \cite{r06}. In this paper, we adopt a simplified version
\cite{r06,r20,r21} of the stationary phase method to perform the
asymptotic analysis. This simplified version has been successfully
applied in the research of underwater communication
\cite{r22,r23,r24}, antenna design \cite{r25,r26}, and NFFF
transformation \cite{r27,r28}.

Theoretically, the LOA can acquired precise results from SIs when
the observation point moves to infinity \cite{r18,r21}. But for
practical usage, an empirical distance is needed to justify its
applicability, which is the NFFF boundary. Although an error
analysis by estimation of higher-order contributions may reveal the
boundary, the analysis becomes very difficult in stratified
structures. Here we compare the asymptotic results with the accurate
numerical ones in different stratified structures. In this way a
universal empirical boundary is obtained. It is found that the NFFF
boundary is mainly affected by lateral waves, which correspond to
the branch point contributions in SIs \cite{r03,r04,r05,r06}.
Besides, the boundary is sensitive to the structure configurations
and the location of the dipole. Moreover, two kinds of treatments in
the asymptotic method are carried out and their numerical results
are compared. The equivalence between the field expressions obtained
from the asymptotic method and reciprocal theorem is also
demonstrated.

The paper is arranged as follow: Section 2 sets our model of the
stratified structures and presents the LOA formalism. Section 3 and
4 discuss the far-field distributions and the NFFF boundary in
bilayer and trilayer structures, respectively. Then the results are
generalized to multilayer structures and proved to be universally
applicable in Sec. 5. Finally, Sec. 6 presents our concluding
remarks.


\section{2. Modeling the stratified structure}

\begin{figure}[h]
\centering\includegraphics[width=8cm]{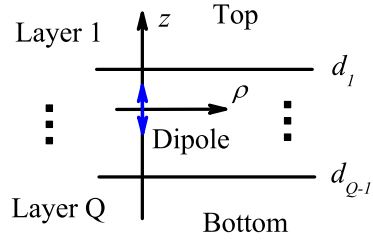} \caption{Schematic of
a stratified structure. The cylindrical coordinates are set and the
origin is always on the VED.}
\end{figure}

The model studied in this paper consists of a lossless layered
structure with the stratification along the $z$ direction. It
contains $Q$ layers and $Q-1$ interfaces from top to bottom, as
depicted in Fig. 1. Since we concentrate on the far field, the two
semispaces, i. e., the top and bottom layers, are focused on and the
intermediate regions are ignored when their details are not needed.
The lower surface of the top layer is at $z=d_{1}$ and the upper
surface of the bottom layer is at $z=d_{Q-1}$, respectively. A
vertical electric dipole (VED) is located on the origin inside the
structure. For convenience, we consider the $z$ component of the
electric fields $E_{z}$ generated by the VED. Other field components
and other dipole orientations will be discussed in Sec. 5. In Fig.
1, the cylindrical coordinates $(\rho, \varphi, z)$ are set. If the
VED is either in the top layer or in the intermediate region, the
field $E_{z}$ in the top and bottom layers is expressed as
\cite{r06}
$$
E_{z}(\rho, z)=\left\{
\begin{array}{rcl}
\displaystyle
[P.F.]\delta_{1q}-\frac{\omega\mu_{0}\mu_{1}j_{z}}{8\pi k_{1}^{2}} \int_{-\infty}^{\infty}dk_{\rho}\frac{k_{\rho}^{3}}{k_{qz}}H_{0}^{(1)} (k_{\rho}\rho)[C_{1}e^{ik_{1z}(z-d_{1})}],       &      & {z>d_{1}.}\\
\displaystyle -\frac{\omega\mu_{0}\mu_{Q}j_{z}}{8\pi
k_{Q}^{2}}\int_{-\infty}^{\infty}
dk_{\rho}\frac{k_{\rho}^{3}}{k_{qz}}H_{0}^{(1)}(k_{\rho}\rho)
[C_{Q}e^{-ik_{Qz}(z-d_{Q-1})}],     &      & {z<d_{Q-1}.}
\end{array} \right.
\eqno{(1)}
$$
The quantities in this expression are as follows. $H_{0}^{(1)}$
represents the Hankel function of the first kind; $j_{z}$ is the
dipole moment in the $z$ direction; $C_{1}$ and $C_{Q}$ are
respectively the scattering coefficients in the top and bottom
layers; $k_{q}=\omega\sqrt{\varepsilon_{q}\mu_{q}}/c
=(k_{\rho}^{2}+k_{qz}^{2})^{1/2}$ is the wave vector in the $q$th
layer; $\omega$ is the angular frequency; $c$ is light speed in
vacuum; $\mu_{0}$ and $\mu_{q}$ are respectively the permeability in
vacuum and in the $q$th layer;  $P.F.$ standing for primary field,
is the $E_{z}$ generated by the VED in a homogeneous background;
$\delta_{1q}$ is Kronecker delta. In Eq. (1), the VED is not allowed
to appear in the bottom layer. However, if the VED is in the bottom
layer, one may use the reversed geometry instead.


\section{3. Bilayer structures}

This section discusses the far-field asymptotic behaviors of dipole
radiations and the NFFF boundary in bilayer structures.

\begin{figure}[h]
\centering\includegraphics[width=7cm]{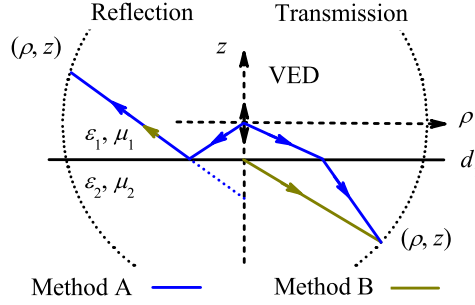} \caption{Physical
meanings of SPs in a bilayer structure. Method A: blue lines; Method
B: yellow lines. For reflections, Method A and B show the same
picture that the reflected field is equal to a field generated by an
image point of VED mirrored by the interface. For transmissions,
Method A describes a refraction that satisfies Snell's law, while
Method B sets the start point on the interface right below the VED.}
\end{figure}

In this case, the intermediated region in Fig. 1 is removed, and the
interfaces at $z=d_{1}$ and $d_{Q-1}$ merge into one, as shown by
Fig. 2. The top and bottom layers are also denoted as layer 1 and 2,
respectively.

\subsection{3.1. The expressions of the $E_{z}$ in the far-field zones}

In the case of bilayer structures shown in Fig. 2 ($d_{1}<0$),
$C_{1}$ and $C_{Q}$ in Eq. (1) are expressed as
$$
\left\{
\begin{array}{rcl}
\displaystyle
C_{1}=r_{12}e^{-ik_{1z}d_{1}}=\frac{k_{1z}/\varepsilon_{1}-k_{2z}/ \varepsilon_{2}}{k_{1z}/\varepsilon_{1}+k_{2z}/\varepsilon_{2}}e^{-ik_{1z}d_{1}},\\
\displaystyle
C_{2}=t_{12}e^{-ik_{1z}d_{1}}=\frac{2k_{1z}/\varepsilon_{1}}
{k_{1z}/\varepsilon_{1}+k_{2z}/\varepsilon_{2}}e^{-ik_{1z}d_{1}},
\end{array} \right.
\eqno{(2)}
$$
where $\varepsilon_{q}$ is the relative permittivity of the $q$th
layer. In this paper, $r_{ij}$ and $t_{ij}$ denote the Fresnel
scattering coefficients of a single interface when light incidents
from layers $i$ to $j$. Here we carry out a detailed derivation on
the transmission field with two asymptotic treatments. In this
process, the similarities and discrepancies between these two
treatments are disclosed.

Substitution of Eq. (2) into (1) and application of the asymptotic
form of Hankel function give the expression of $E_{z}$ in layer 2:
$$
E_{z}(z<d_{1})=-\frac{\omega\mu_{0}\mu_{2}j_{z}}{8\pi k_{2}^{2}}
\int_{-\infty}^{\infty}dk_{\rho}\frac{k_{\rho}^{3}}{k_{1z}}t_{12}
e^{i(k_{\rho}\rho-\pi/4)}e^{-ik_{2z}(z-d_{1})}e^{-ik_{1z}d_{1}}.
\eqno{(3)}
$$
The SP is obtained by taking derivative of the phase as follows:
$$
\rho+\frac{k_{\rho s}}{k_{1zs}}d_{1}+\frac{k_{\rho
s}}{k_{2zs}}(z-d_{1})=0. \eqno{(4a)}
$$
This way determining the SP is called Method A. Its physical meaning
can be explained by ray theory. On the other hand, if the factor
$\exp(-ik_{1z}d_{1})$ in Eq. (3) is a slowly varying one, it can be
taken out of the integral, and the remaining phase gives the SP as
follow:
$$
\rho+\frac{k_{\rho s}}{k_{2zs}}(z-d_{1})=0. \eqno{(4b)}
$$
This way is called Method B \cite{r26,r27}. In Eqs. (4a) and (4b),
the subscript ``s'' stands for stationary points. Here we emphasize
the features of these two methods. Method A treats
$\exp(-ik_{1z}d_{1})$ as an oscillatory factor. As a result, the
term $k_{\rho s}d_{1}/k_{1zs}$ is included in Eq. (4a) in
determining the SP. It describes the wave paths in both layers and
gives the refraction picture shown by the blue lines in Fig. 2. By
contrast, Method B treats $\exp(-ik_{1z}d_{1})$ as a slowly varying
one so that the term $k_{\rho s}d_{1}/k_{1zs}$ is excluded in Eq.
(4b) in determining the SP. This description is merely applicable to
the wave path in layer 2, as shown by the yellow lines in Fig. 2.

Under these points of view, Method A gives the electric field as
$$
E_{z}(z<d_{1})=-\frac{\omega\mu_{0}\mu_{2}j_{z}}{8\pi
k_{2}^{2}}t_{12} (k_{\rho s})\frac{k_{2zs}}{k_{1zs}}k_{\rho
s}^{2}\int_{-\infty}^{\infty}
dk_{\rho}\frac{k_{\rho}}{k_{2z}}H_{0}^{(1)}(k_{\rho}\rho)
e^{-ik_{2z}(z-d_{1})-ik_{1z}d_{1}}, \eqno{(5a)}
$$
while Method B gives
$$
E_{z}(z<d_{1})=-\frac{\omega\mu_{0}\mu_{2}j_{z}}{8\pi
k_{2}^{2}}t_{12} (k_{\rho s})\frac{k_{2zs}}{k_{1zs}}k_{\rho
s}^{2}e^{-ik_{1zs}d_{1}}
\int_{-\infty}^{\infty}dk_{\rho}\frac{k_{\rho}}{k_{2z}}H_{0}^{(1)}
(k_{\rho}\rho)e^{-ik_{2z}|z-d_{1}|}. \eqno{(5b)}
$$
The simplification in Method B is that the integral in Eq. (5b) can
be expressed as spherical waves using the Sommerfeld identity
\cite{r06,r20}
$$
\frac{e^{ikr}}{r}=\frac{i}{2}\int_{-\infty}^{\infty}dk_{\rho}\frac{k_{\rho}}
{k_{z}}H_{0}^{(1)}(k_{\rho}\rho)e^{-ik_{z}|z|}.
$$
By contrast, Eq. (5a) cannot be further reduced in mathematics
because its exponential term contains two different wave vectors. We
have to carry out the integral according to Eq. (4a). The
corresponding ray interpretation is demonstrated by blue lines in
Fig. 2.

After integrations, the asymptotic expressions of $E_{z}$ in Method
A and B are respectively given by
$$
\left\{
\begin{array}{rcl}
E_{z}(z<d_{1}) & = &\displaystyle -\frac{\omega\mu_{0}\mu_{2}j_{z}}{8\pi k_{2}^{2}} t_{12} (k_{\rho s})\frac{k_{2zs}}{k_{1zs}}k_{\rho s}^{2}\left[\frac{2}{i} \frac{e^{ik_{1}\triangle r_{1}}e^{ik_{2}\triangle r_{2}}}{\triangle r_{1}+ \triangle r_{2}}\right]\\
& = &\displaystyle t_{12}(k_{\rho s})\frac{k_{2zs}}{k_{1zs}}e^{i(k_{1}-k_{2})\triangle r_{1}}\left\{\frac{i\omega\mu_{0}\mu_{2}j_{z}}{4\pi k_{2}^{2}}k_{\rho s}^{2} \frac{e^{ik_{2}(\triangle r_{1}+\triangle r_{2})}}{\triangle r_{1}+ \triangle r_{2}}\right\}\\
\triangle r_{1} & = &\displaystyle \frac{k_{1}}{k_{1zs}}|d_{1}|\\
\triangle r_{2} & = &\displaystyle \frac{k_{2}}{k_{2zs}}|z-d_{1}|
\end{array} \right.
\eqno{(6a)}
$$
and
$$
\left\{
\begin{array}{rcl}
E_{z}(z<d_{1}) & = &\displaystyle -\frac{\omega\mu_{0}\mu_{2}j_{z}}{8\pi k_{2}^{2}}t_{12} (k_{\rho s})\frac{k_{2zs}}{k_{1zs}}k_{\rho s}^{2}e^{-ik_{1zs}d_{1}} \left[\frac{2}{i}\frac{e^{ik_{2}r_{2}}}{r_{2}}\right]\\
& = &\displaystyle t_{12}(k_{\rho s})\frac{k_{2zs}}{k_{1zs}}e^{-ik_{1zs}d_{1}} \left\{\frac{i\omega\mu_{0}\mu_{2}j_{z}}{4\pi k_{2}^{2}}k_{\rho s}^{2} \frac{e^{ik_{2}r_{2}}}{r_{2}}\right\}\\
r_{2} & = & \sqrt{\rho^{2}+(z-d_{1})^{2}}
\end{array}. \right.
\eqno{(6b)}
$$
The expressions in the square brackets in Eqs. (6a) and (6b) are the
resultants of the integrals in Eqs. (5a) and (5b), and those in the
curly brackets represent the propagations of $E_{z}$ along the blue
and yellow lines shown in Fig. 2, respectively. Thus it is easy to
see that Eq. (6a) considers $E_{z}$ as spherical waves in both
layers and describes a refraction process at the interface. The
factor $\exp[i(k_{1}-k_{2})\triangle r_{1}]$ in Eq. (6a) is the
phase correction for the spherical wave propagating in layer 1, as
can be seen by comparison of Eqs. (6a) and (6b). The term
$t_{12}(k_{\rho s})k_{2zs}/k_{1zs}$ arises from the boundary
condition.

The physical explanation of Eq. (6b) is similar. The factor
$\exp(-ik_{1z}d_{1})$ reflects the influence of the wave in layer 1.
In layer 1, $E_{z}$ is a plane wave propagating along the $z$
direction, and after crossing the interface at the point right below
the VED, it turns to the spherical wave and propagates to the
observation points.

Method B was given by Refs. \cite{r26,r27}. Comparatively, Method A
provides a more accurate description and asymptotic expression for
the transmission than Method B. That is why we suggest Method A in
this paper. However, the discrepancies of these two methods tend to
be trivial as $d_{1}/r$ goes to zero. Another treatment of the
transmission field \cite{r22,r23,r24} employed Eqs. (4a) and (6b).
In other words, it considered $\exp(-ik_{1z}d_{1})$ first as an
oscillatory term to calculate the SP, and then as a slowly varying
one to take out the integral. The accuracy of this method is between
Method A and B. We do not discuss it in the following.

Now let us show the equivalence between the asymptotic method,
Method B, and the reciprocal theorem. Eq. (6b) can be reformed as
$$
E_{z}(z<d_{1})=\frac{\varepsilon_{2}}{\varepsilon_{1}}t_{21}(k_{\rho
s})e^{-ik_{1zs}d_{1}}\left\{\frac{i\omega\mu_{0}\mu_{2}j_{z}}{4\pi
k_{2}^{2}}k_{\rho s}^{2} \frac{e^{ik_{2}r_{2}}}{r_{2}}\right\}
\eqno{(7)}
$$
because
$$
t_{12}(k_{\rho s})\frac{k_{2zs}}{k_{1zs}}=\frac{\varepsilon_{2}}
{\varepsilon_{1}}\frac{2k_{2zs}/\varepsilon_{2}}{k_{1zs}/\varepsilon_{1}
+k_{2zs}/\varepsilon_{2}}=\frac{\varepsilon_{2}}{\varepsilon_{1}}
t_{21}(k_{\rho s}).
$$
Since $t_{21}(k_{\rho s})$ is the transmission coefficient when
light incidents from layers 2 to 1, Eq. (7) describes the reversal
propagation picture: the $E_{z}$ is now generated by a VED at
far-field zone and propagates back alone the reversal direction of
the yellow line in Fig. 2. After crossing the interface, the
spherical wave turns to a plane wave and propagates to the origin.
That the spherical wave emitted by a test source turns to a plane
wave after passing the interface is the treatment of the reciprocal
theorem. In Ref. \cite{r32}, the turning point was chosen such that
it was at the line along the direction of the VED. Therefore, the
result there was the same as Eq. (6b) here. The conclusion is that
Method B is equivalent to the reciprocal theorem.

When the observation points and the source are in the same layer,
the reflection process described by Methods A and B are identical
\cite{r02,r03,r04,r05,r06}, as shown in Fig. 2. The asymptotic
expression of $E_{z}$ in layer 1 is expressed as
$$
\left\{
\begin{array}{rcl}
E_{z}(z>d_{1}) & = &\displaystyle [P.F.]\delta_{1q}+r_{12}(k_{\rho s}) \left\{\frac{i\omega\mu_{0}\mu_{1}j_{z}}{4\pi k_{1}^{2}}k_{\rho s}^{2} \frac{e^{ik_{1}r_{1}}}{r_{1}}\right\}\\
r_{1} & = & \sqrt{\rho^{2}+(z-2d_{1})^{2}}
\end{array}. \right.
\eqno{(8)}
$$

\subsection{3.2. The features of the fields}

With the formulas derived above, we are able to do numerical
computation. The numerical results of the asymptotic methods are
compared to the precise numerical ones \cite{r10}. This enables us
to acquire the NFFF boundary. As an example, we take
$\lambda_{0}=1.5\mu m$ and $\mu_{q}=1$. As can be seen later, our
conclusions are valid for other wavelengths and permeabilities.
Since $E_{z}$ is of an axial symmetry, the angular direction
$\varphi$ is not considered. Instead, the observation points are
labeled by the spherical coordinates $(r, \theta, \varphi)$. Figure
3 shows the value of $|E_{z}|$ as a function of angle $\theta$, at
different distance $r$. When $0\leq \theta\leq \pi/2$, the
observation point is in layer 1, while when $\pi/2\leq \theta\leq
\pi$, the observation point is in layer 2. In Fig. 3, the dielectric
constants in layers 1 and 2 are set to be 1 and 2.25, respectively.
So, the upper and lower layers are called lower and higher
refraction index (RI) layers, respectively.

\begin{figure}[h]
\centering\includegraphics[width=8cm]{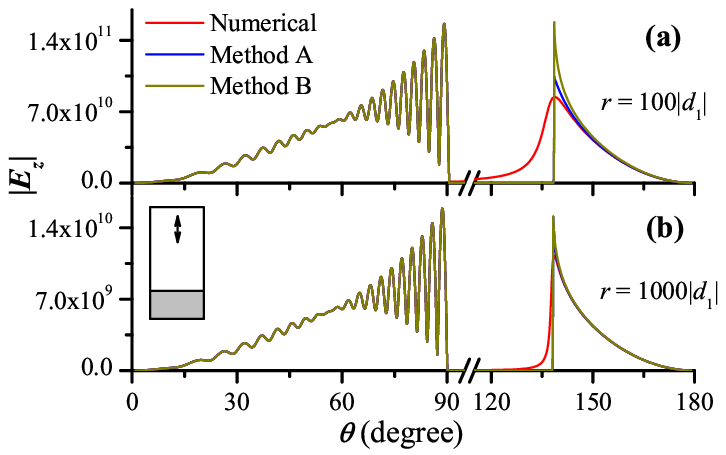}
\centering\includegraphics[width=8cm]{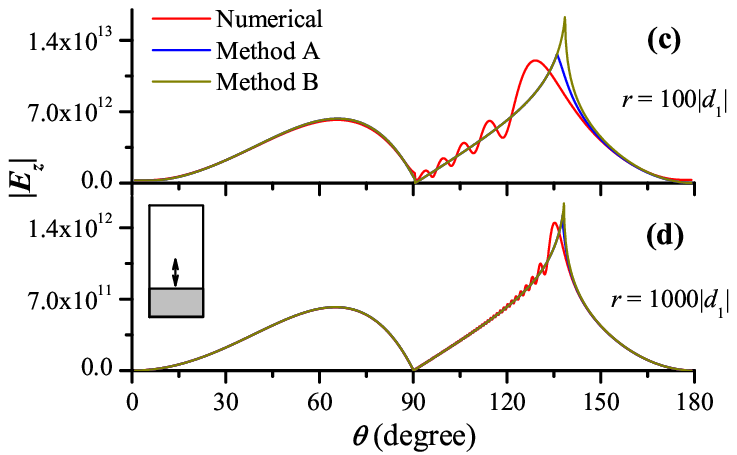} \caption{$|E_{z}|$
as a function of angle $\theta$ for different distance $r$.
Structural schematics are illustrated in the insets. The critical
angle is
$\theta_{c}\sim$($180^{\circ}$--$42^{\circ}$)=$138^{\circ}$. (a)
($|d_{1}|, r$)=($10, 1000$)$\lambda_{0}$. (b) ($|d_{1}|, r$)=($10,
10000$)$\lambda_{0}$. (c) ($|d_{1}|, r$)=($0.1, 10$)$\lambda_{0}$.
(d) ($|d_{1}|, r$)=($0.1, 100$)$\lambda_{0}$.}
\end{figure}

In Fig. 3, the precise numerical results \cite{r10} are plotted by
red lines for comparison. For reflections, the oscillation curves in
Figs. 3(a) and 3(b) reflect the interference between the primary
field and reflected field, while this pattern is not shown in Figs.
3(c) and 3(d) because the distance between the VED and its image is
too close. The asymptotic results fit the numerical ones very well
when $r$ is of the order of $10\lambda_{0}$, or smaller
$r\sim\lambda_{0}$ \cite{r23}, which agrees with the NFFF boundary
in a vacuum background.

When transmitting from the lower RI layer to the higher RI one,
there exists a critical angle $\theta_{c}$ above and below which are
allowed and forbidden regions, respectively \cite{r36,r37}. The
boundaries of the forbidden regions are indicated by the
transmission peaks in Fig. 3. Please note that $\theta_{c}$ is
related to $d_{1}$ value, so that the transmission peaks in the four
panels of Fig. 3 have slight differences. In Figs. 3(c) and 3(d),
the curves in the forbidden regions show ripples which is an
interference effect caused by the boundary continuity \cite{r35}.
Since the ripple is a near field effect, it does show up in Figs.
3(a) and 3(b) where the observation distance is far. When the
distance is less, the ripple will become evident. Moreover, when the
real part of $k_{1z}$ vanishes, $\exp(-ik_{1z}d_{1})$ should be
treated as a slowly varying term, which means that Method A can be
replaced by Method B near $\theta_{c}$ and within the forbidden
region.

In the allowed region, although Method A provides more accurate
results, both of them coincide with the numerical results. However,
within the forbidden region and around $\theta_{c}$, there is a big
difference between our results and the numerical ones. When the VED
is far away from the interface, as shown in Fig. 3(a), the
asymptotic result decays rapidly around $\theta_{c}$ due to the
effect of $\exp(-ik_{1z}d_{1})$, while the red line shows a slow
attenuation. At first glance, this seems strange since the numerical
results also contain $\exp(-ik_{1z}d_{1})$. This can be answered by
Eqs. (3) and (6b). The numerical method treats
$\exp(-ik_{1z}d_{1})/k_{1z}$ in Eq. (3) as a whole factor, which
means that when both of the numerator and denominator tend to 0,
$k_{1z}$ slows down the decay of $\exp(-ik_{1z}d_{1})$. While the
asymptotic methods treat $1/k_{1z}$ as a constant, implying that
only the decay of $\exp(-ik_{1z}d_{1})$ is considered. In SIs,
$k_{1z}=0$ corresponds to the $k_{1}$ branch point, and at this
point the integration gives a lateral wave
\cite{r02,r03,r04,r05,r06}. Lateral waves are the surface waves that
mainly exist in the higher RI semispace and exponentially decay in
the lower RI one. This explanation is also in line with the
description of the steepest descent method. Near $\theta_{c}$, the
steepest descent path approaches the $k_{1}$ branch point and its
contribution to the integral becomes inegligible. In the forbidden
region, the branch point is enclosed within the deformed integration
path, and causes the lateral waves to appear in the forbidden
region. Since the lateral waves represent a higher-order
attenuation, it is not covered by the LOA. This indicates that the
differences between the asymptotic results and numerical results
come from the branch point contributions.

\begin{figure}[h]
\centering\includegraphics[width=8cm]{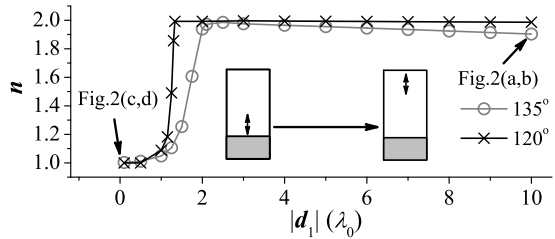} \caption{Results of
the power series fitting the amplitude of the lateral waves at
angles $120^{\circ}$ and $135^{\circ}$ in Fig. 3. Fitting functioin
is $|E_{z}|$=$ar^{-n}$, where $a$ (not show here) and $n$ are
fitting coefficients. The fitting range is
$r$=$4000\sim10000\lambda_{0}$ \cite{r38}.}
\end{figure}

The branch point contributions can also be treated by asymptotic
methods, e. g., branch cut integrals based on the steepest descent
method \cite{r29,r30,r31}, or uniform asymptotic expansions
\cite{r06}. However, these methods regard $\exp(-ik_{1z}d_{1})$ as a
slowly varying term \cite{r06,r29,r30}, and numerical computation is
resorted \cite{r13,r15,r31}. It is difficult to generalize such
asymptotic expressions to the cases of multilayer structures
\cite{r30}. The exploration of asymptotic expressions for branch cut
integrals is beyond the scope of this paper, so here we numerically
calculate the decay behaviors of the lateral waves at angles
$120^{\circ}$ and $135^{\circ}$ in Fig. 3, and show them in Fig. 4.
It is seen from Fig. 4 that when $|d_{1}|>2\lambda_{0}$, the lateral
waves decay as $r^{-2}$. This coincides with the case shown in Figs.
3(a) and 3(b) where $r$ decreases by one order of magnitude while
the field decreases by two orders. When
$\lambda_{0}<|d_{1}|<2\lambda_{0}$, the way of the decay of the
lateral waves turns to $r^{-1}$, so that in Figs. 3(c) and 3(d) the
numerical results and the asymptotic results have similar
distributions in the forbidden region.

\subsection{3.3. Determining the NFFF boundary}

Having the knowledge of the decay behaviors of the lateral waves, we
discuss the NFFF boundary in the bottom layer. As can be seen from
Fig. 3, the boundary may be chosen as $L\sim1000|d_{1}|$, which
satisfies the evaluation of the order of magnitude indicated by Fig.
4. However, this distance should be modified because of two reasons.
One is that for a very small $|d_{1}|$, e. g.,
$|d_{1}|=0.001\lambda_{0}$, $L\sim\lambda_{0}$ may not eliminate the
influence of the lateral waves (numerically verified), which means
that a lower boundary should be better. Considering Fig. 3(d), the
boundary may be written as $L\sim1000\times$Max$\{|d_{1}|,
0.1\lambda_{0}\}$, where Max$\{,\}$ means ``choose the larger one''.
The other is that compared to the one of a homogeneous background,
the NFFF boundary here has a looser relationship with $\lambda_{0}$.
Although the two $|d_{1}|$ values in Fig. 3 differ by 100 times, the
NFFF boundary only depends on $|d_{1}|$. Moreover, as implied by
Eqs. (3), (6a) and (6b), the field distribution has a scale
invariance over $\lambda_{0}$ in a non-dispersive media, indicating
that the field behaviors shown in Fig. 3 stand also for other light
wavelengths. So, what influences does $\lambda_{0}$ have on the NFFF
boundary? Except for the lower boundary, it mainly decides the
location of the forbidden region and introduces dispersion. The
dispersion considered, the NFFF boundary is suggested to be
$$
L\sim1000\frac{n_{max}}{n_{min}}\times\mbox{Max}\{D,
0.1\lambda_{0}\}, \eqno{(9)}
$$
where $n_{max}$ and $n_{min}$ respectively represent the largest and
smallest RI in the structure,
$n_{q}=(\varepsilon_{q}\mu_{q})^{1/2}$. Equation (9) is the most
significant conclusion of this paper, and it is valid for any
stratified configuration, as will be illustrated below. Here, a
dimensional parameter $D$ is introduced. In the present bilayer
structure, $D$ equals to $|d_{1}|$. For other configurations with
more layers, $D$ needs to be modified, as will be shown below.

In the following, we test Eq. (9) with other configurations. To
avoid repetition, the subsequent discussion will not compare the
results of different observation distances, but only focus on those
at $r=100\lambda_{0}$, a distance less than that given by Eq. (9).

\begin{figure}[h]
\centering\includegraphics[width=8cm]{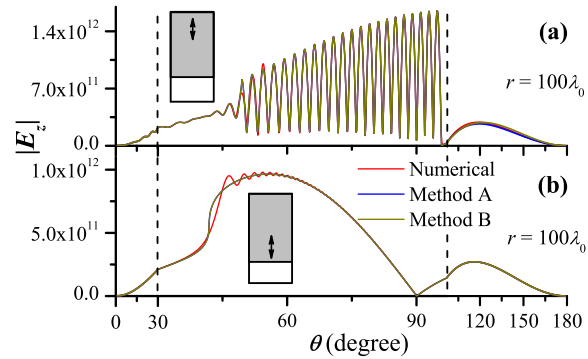} \caption{$|E_{z}|$ as
a function of angle $\theta$ at distance $r$=$100\lambda_{0}$.
Structural schematics are illustrated in the insets. The critical
angle is $\theta_{c}\sim42^{\circ}$. (a) $|d_{1}|$=$10\lambda_{0}$.
(b) $|d_{1}|$=$0.1\lambda_{0}$. The coordinate between
$30^{\circ}\sim96^{\circ}$ is stretched to clearly show the
oscillation details.}
\end{figure}

Now we set the dielectric constants in layers 1 and 2 to be 2.25 and
1, respectively. So, in this case the VED is in a higher RI layer.
$|E_{z}|$ as a function of angle $\theta$ for two $|d_{1}|$ values
is plotted in Fig. 5. In the range $0\leq\theta\leq\pi/2$, i. e., in
layer 1, the oscillations shown in Figs. 5(a) and 5(b) reflect the
interference between the primary and reflected fields as have been
mentioned in analyzing Fig. 3. Since the lateral waves favor to
appear in the higher RI semispace, in the present case, they mainly
influence the reflection.

In Fig. 5, the coordinates between $30^{\circ}\sim96^{\circ}$ are
stretched to illustrate the oscillations. Around $\theta_{c}$, the
asymptotic results and the numerical results have slight differences
in the range $45^{\circ}\sim50^{\circ}$, so that Eq. (9) is also
applicable here. As for the transmission, since now light is from
the layer with lower RI to that with higher RI, no forbidden regions
appear in this configuration. Light refracts and the asymptotic
results fit the numerical results quite well.

According to the discussions above, it is known that the NFFF
boundary is mainly affected by the lateral waves. In the higher RI
semispace, the boundary is given by Eq. (9) and much larger than the
operating wavelength; while in the lower RI semispace, the boundary
is about ten wavelengths. Please notice that it is not appropriate
to define the NFFF boundary by distinguishing the allowed and
forbidden region, because there is no forbidden region in the
configurations shown in Fig. 5.


\section{4. Trilayer structures}

Now we investigate the case of trilayer structure. The three layers
are called the top, middle and bottom layers, or layers 1, 2 and 3,
as illustrated in Fig, 6. Two cases are involved where the location
of the VED is in the top and middle layers, respectively.

Before presenting numerical results, three illustrations ought to be
addressed. Firstly, different from bilayer structures, there will
occur in a trilayer structure three kinds of surface modes
\cite{r02,r03,r04,r05,r06,r39,r40}: guided modes, proper complex
modes and improper complex modes, which correspond to poles in SIs.
The first two belong to normal modes that build up the point
spectrum in eigenvalue problems. While the last one is commonly
referred as leaky modes which violates the radiation condition and
only shows up in a certain range of angle in space \cite{r39,r40}.
The guided modes occur in the middle layer of a mode guiding
configuration where the middle layer possesses the largest RI. The
proper complex modes cannot propagate far from the source due to
their decaying nature. Therefore, these two kinds of modes will not
be touched in the following discussions. The leaky modes affect the
far-field distribution and will be considered. Secondly, because
there are two interfaces in this structure, light multi-reflects in
the middle layer and may cause more oscillations in the far-field
distribution. We will not explain all the formations of these
oscillations, but focus on the ones that may affect the NFFF
boundary. Moreover, the differences between Method A and B are not
obvious in the previous section because $d_{1}/r$ are small, but
will be magnified by the multi-reflection in the middle layer.
Thirdly, there are six possible configurations of dielectric
constant distribution for a trilayer structure. We merely study two
of them: $[\varepsilon: 1-2.25-12]$ and $[\varepsilon: 2.25-12-1]$,
which are often encountered in experiments. Our conclusions are
easily extended to other configurations.

\subsection{4.1. The VED is in the top layer}

\begin{figure}[h]
\centering\includegraphics[width=7cm]{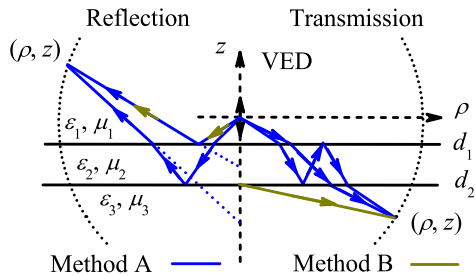} \caption{Physical
meanings of SPs when a VED is located in the top layer. Method A:
blue lines; Method B: yellow lines. Concerning reflections, Method A
shows that the multi-reflection process in the middle layer
contributes to the reflection in layer 1, while Method B only
considers the direct reflection from the interface at $d_{1}$. For
transmissions, Method A shows that the multi-reflection process
contributes to the transmission in layer 3, while Method B sets the
start point on the interface at $d_{2}$ right below the VED. In the
figure, only the modes $m$=0 and 1 in the multi-reflection are
depicted.}
\end{figure}

When the VED is in the top layer, as shown in Fig. 6
($d_{2}<d_{1}<0$), the scattering coefficients $C_{1}$ and $C_{Q}$
are expressed as
$$
\left\{
\begin{array}{lcl}
\displaystyle
C_{1}=r_{12}+\frac{t_{12}r_{23}t_{21}e^{2ik_{2z}(d_{1}-d_{2})}} {1-r_{21}r_{23}e^{2ik_{2z}(d_{1}-d_{2})}}e^{-ik_{1z}d_{1}},\\
\displaystyle C_{3}=\frac{t_{12}t_{23}e^{2ik_{2z}(d_{1}-d_{2})}}
{1-r_{21}r_{23}e^{2ik_{2z}(d_{1}-d_{2})}}e^{-ik_{1z}d_{1}},
\end{array} \right.
\eqno{(10)}
$$
where the expressions of the coefficients $r_{ij}$ and $t_{ij}$ can
be found in Eq. (2). After expanding Eq. (10) with the geometrical
optics series and following the steps in Sec. 3, the SPs of Method A
for reflections can be obtained by,
$$
\rho-\frac{k_{\rho s}^{(m)}}{k_{1zs}^{(m)}}(z-2d_{1})-\frac{k_{\rho
s}^{(m)}}{k_{2zs}^{(m)}}2m(d_{1}-d_{2})=0, \eqno{(11)}
$$
where $m\in[0,\infty)$ is the order of the geometrical optics
series. The SP of Method B is obtained by letting $m=0$ in Eq. (11).
In the same way, the SPs of Method A for transmissions meet
$$
\rho+\frac{k_{\rho s}^{(m)}}{k_{1zs}^{(m)}}d_{1}-\frac{k_{\rho
s}^{(m)}}{k_{2zs}^{(m)}}(2m+1)(d_{1}-d_{2})+\frac{k_{\rho
s}^{(m)}}{k_{3zs}^{(m)}}(z-d_{2})=0. \eqno{(12)}
$$
Since Method B only consider the phase variations in the observation
layer, its transmission SP has the same form as Eq. (4b), with
$d_{1}$ and $k_{2zs}$ being replaced by $d_{2}$ and $k_{3zs}$
respectively. The physical meanings of the SPs are depicted in Fig.
6. Method A describes a multi-reflection process, so that its
$E_{z}$ expressions for reflections and transmissions are in the
form of series expansions. Each term of the expressions is similar
to Eq. (6a), where $\triangle r_{2}$ now represents the
multi-reflection path in the middle layer and it is also needed for
describing the path in layer 3. As for Method B, the asymptotic
expressions are similar to Eqs. (6b) and (8). Moreover, the
equivalence between the expressions of Method B and reciprocal
theorem can also be proven, in a similar way to that done in Eq.
(7).

\begin{figure}[h]
\centering\includegraphics[width=8cm]{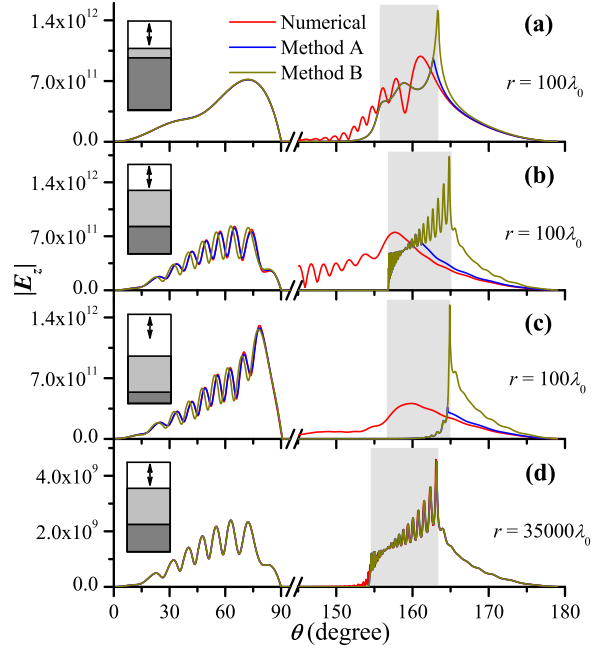} \caption{$|E_{z}|$
as a function of angle $\theta$ for four geometry configurations
with [$\varepsilon: 1-2.25-12$]. Two critical angles are
$\theta_{c1}\sim$($180^{\circ}- 17^{\circ}$)=$163^{\circ}$ and
$\theta_{c2}\sim$($180^{\circ}- 26^{\circ}$)=$154^{\circ}$.
Structural schematics are illustrated in the insets. (a) ($|d_{1}|,
|d_{2}|, r$)=($0.1, 1.1, 100$)$\lambda_{0}$. (b) ($|d_{1}|, |d_{2}|,
r$)=($0.1, 10.1, 100$)$\lambda_{0}$. (c) ($|d_{1}|, |d_{2}|,
r$)=($1.1, 10.1, 100$)$\lambda_{0}$. (d) ($|d_{1}|, |d_{2}|,
r$)=($0.1, 10.1, 35000$)$\lambda_{0}$. The shaded areas between
$\theta_{c1}$ and $\theta_{c2}$ are the second total internal
reflection regions.}
\end{figure}

Figure 7 plots the $E_{z}$ distributions in the case of
$[\varepsilon: 1-2.25-12]$ configuration. Since layers 2 and 3 have
higher RI, there are two critical angles:
$\theta_{c1}\sim\pi-asin(1/\sqrt{12})$  and
$\theta_{c1}\sim\pi-asin(1/\sqrt{2.25})$. $\theta_{c1}$ determines
the forbidden region. The range between $\theta_{c1}$ and
$\theta_{c2}$ is shaded in Fig. 7 and is termed as the second total
internal reflection region.

In Fig. 7(a), the thickness of the middle layer is $\lambda_{0}$. It
is seen that the differences between the asymptotic results and
numerical results emerge in the transmission field. These
differences are caused by the lateral waves as well. The dimensional
parameter $D$ in Eq. (9) now should equal to the distance between
VED and the lower interface at $z=d_{2}$. The calculated NFFF
boundary is approximately 35 times of the observation distance,
which is enough to eliminate the influence of lateral waves
(numerical verified, not show here).

In Fig. 7(b), the middle layer is thicker: its thickness is
$10\lambda_{0}$. This figure shows two significant features. One is
that Method A and B have noticeable differences in both of the
reflections and transmissions. Because Method A takes into account
the multi-reflection, its results are closer to the numerical ones.
The other is that in the second total internal reflection region,
the asymptotic curve shows rapid oscillation [The coefficients
$C_{1}$ and $C_{3}$ in Eq. (10) are of the form of multi-reflection,
and in this sense, there is also multi-reflection in Method B]. The
formation of the oscillation can be explained as following: although
the light decays in the top layer, part of the energy can be
transferred into the middle layer through evanescent waves and
become propagation ones. Then in the middle layer it scatters at the
lower boundary, and has a total internal reflection at upper
boundary, which further forms a multi-reflection to yield the
oscillation. When the VED moves away from the interface, the
oscillation will die away. This is verified by Fig. 7(c) where
$|d_{1}|$ changes from $0.1\lambda_{0}$ to $\lambda_{0}$. It is seen
from Fig. 7(c) that the oscillations disappear and the far-field
distribution decreases. It is worth mentioning that the description
here is the picture of leaky waves. A comparatively thicker middle
layer allows many leaky modes to exist, resulting in the rapid
oscillation in the far-field distribution. As a test of the NFFF
boundary, Fig. 7(d) plots the field distribution of the same
configuration in Fig. 7(b) but with the observation point being at
the NFFF boundary, i. e., with the observation distance computed by
Eq. (9). The results fit each other very well.

\begin{figure}[h]
\centering\includegraphics[width=8cm]{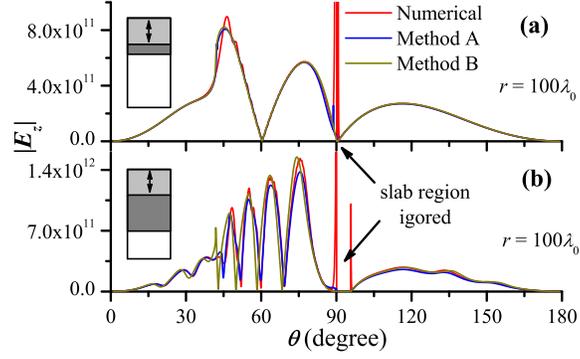} \caption{$|E_{z}|$ as
a function of angle $\theta$ at distance $r$=$100\lambda_{0}$ for
two geometry configurations with [$\varepsilon: 2.25-12-1$]. The two
critical angles are $\theta_{c1}\sim26^{\circ}$ and
$\theta_{c2}\sim42^{\circ}$. Structural schematics are illustrated
in the insets. $|d_{1}|$=$0.1\lambda_{0}$. (a)
$|d_{2}|$=$1.1\lambda_{0}$. (b) $|d_{2}|$=$10.1\lambda_{0}$. The
fields in the middle layer are not presented.}
\end{figure}

Fig. 8 shows the $|E_{z}|$ distributions in $[\varepsilon:
2.25-12-1]$ configuration. Since the middle layer possesses the
largest RI, the structure supports the guided modes. The peaks
around $90^{\circ}$ of the numerical results are the decays of the
guide modes in layers 1 and 3. Moreover, there is no forbidden
region in the structure. Consequently, the asymptotic result and the
numerical results fit very well. For reflections, these two kinds of
results show some differences around $\theta_{c2}$ due to the effect
of lateral waves, but they fit each other at the NFFF boundary given
by Eq. (9). Around $90^{\circ}$, the guide modes decay in the lower
RI layers, but they are not covered in the discussion here because a
guided mode is a bounded state that is equivalence to cylindrical
wave and have a decay rate as $1/\sqrt{\rho}$, which means that its
field is much larger than the results of LOA .

\subsection{4.2. The VED is in the middle layer}

\begin{figure}[h]
\centering\includegraphics[width=7cm]{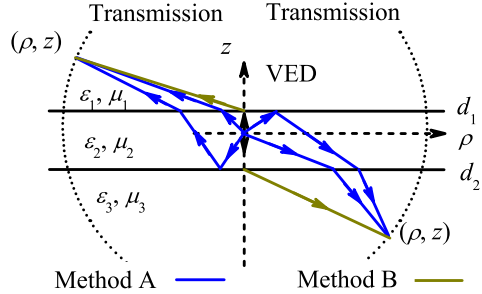} \caption{Physical
meanings of SPs when a VED is located in the middle layer. Method A:
blue lines; Method B: yellow lines. For transmissions in the top
layer, two cases, both involving multi-reflection processes, are
distinguished in Method A, where the propagation direction of the
light is upward or downward after emitted from the VED. For a clear
illustration, only the mode $m$=0 is demonstrated here. Method B
sets the start point on the top interface right above the dipole.
The transmissions in the bottom layer are similar to that in the top
layer, but with the transmission direction reversed.}
\end{figure}

Next, we discuss the case when the VED is located in the middle
layer ($d_{2}<0<d_{1}$), as shown in Fig. 9. For this configuration,
the scattering coefficients $C_{1}$ and $C_{Q}$ are expressed as
$$
\left\{
\begin{array}{lcl}
\displaystyle
C_{1}=t_{21}\frac{e^{ik_{2z}d_{1}}}{1-r_{21}r_{23}e^{2ik_{2z}(d_{1}-d_{2})}} +t_{21}\frac{r_{23}e^{ik_{2z}(d_{1}-2d_{2})}} {1-r_{21}r_{23}e^{2ik_{2z}(d_{1}-d_{2})}},\\
\displaystyle
C_{3}=t_{23}\frac{e^{-ik_{1z}d_{2}}}{1-r_{21}r_{23}e^{2ik_{2z}(d_{1}-d_{2})}}
+t_{23}\frac{r_{21}e^{ik_{2z}(2d_{1}-d_{2})}}
{1-r_{21}r_{23}e^{2ik_{2z}(d_{1}-d_{2})}}.
\end{array} \right.
\eqno{(13)}
$$
For the $|E_{z}|$ in the top layer, Method A expands the two terms
in $C_{1}$ using geometric optics series, which is then substituted
in Eq. (1) to get the SPs:
$$
\left\{
\begin{array}{lcl}
\displaystyle
\rho-\frac{k_{\rho s}^{(m)}}{k_{1zs}^{(m)}}(z-d_{1})-\frac{k_{\rho s}^{(m)}}{k_{2zs}^{(m)}}[2m(d_{1}-d_{2})+d_{1}]=0,\\
\displaystyle \rho-\frac{k_{\rho
s}^{(m)}}{k_{1zs}^{(m)}}(z-d_{1})-\frac{k_{\rho
s}^{(m)}}{k_{2zs}^{(m)}}[2(m+1)(d_{1}-d_{2})-d_{2}]=0.
\end{array} \right.
\eqno{(14)}
$$
On the whole, Eq. (14) represents the light transmitting to the top
layer after multi-reflections. The first equation represents the
case of light propagating upward after being emitted from the VED.
The second equation indicates the fact that light propagates
downwards first, and after is reflected by the lower interface, it
then propagates upwards. The intuitive picture is illustrated in
Fig. 9. This course is also reflected by the expression of $C_{1}$
in Eq. (13). Concerning the ray interpretation of SPs, we can write
the asymptotic expressions similar to Eq. (6a). The transmission
picture of Method B is the same as the above ones, and its
asymptotic expressions are similar to Eq. (6b). The transmission to
the bottom layer is similar to that to the top layer, but with the
direction in reverse.

\begin{figure}[h]
\centering\includegraphics[width=8cm]{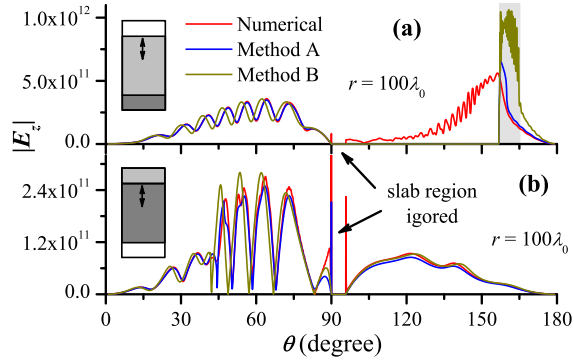} \caption{$|E_{z}|$ as
a function of angle $\theta$ at distance $r$=$100\lambda_{0}$ for
two configurations. The two critical angles are
$\theta_{c1}\sim$($180^{\circ}-17^{\circ}$)=$163^{\circ}$ and
$\theta_{c1}\sim$($180^{\circ}-26^{\circ}$)=$154^{\circ}$.
Structural schematics are illustrated in the insets. ($|d_{1}|,
|d_{2}|,r$)=($0.1, 10, 100)\lambda_{0}$. (a) [$\varepsilon:
1-2.25-12$]. (b) [$\varepsilon: 2.25-12-1$]. The gray area in (a)
represents the second total internal reflection region.}
\end{figure}

The $|E_{z}|$ distributions when the VED is in the middle layer of
trilayer structures are shown in Fig. 10. Figures 10(a) and 7(b)
have similar far-field distributions, and Figs. 10(b) and 8(b) as
well. Therefore, the discussions there are valid for Fig. 10. The
correctness of Eq. (9) is again verified, where the parameter $D$ is
now the thickness of the middle layer. Moreover, Fig. 10 also shows
that Method A provides more accurate results than Method B.

It is known from the discussions above that conclusions about the
NFFF boundary in a trilayer structure are similar to that obtained
in a bilayer structure. The boundary in a higher RI layer is mainly
determined by the lateral waves and satisfies Eq. (9), while the
boundary in the lowest RI layer is about ten wavelengths.


\section{5. General configurations}

In this section, we generalize our conclusions to multilayer
structures, and verify the universality of Eq. (9).

For multilayer structures, light multi-reflects in each layer in the
intermediate region. Consequently, the expressions of the field will
have a recursive fashion, which will make the ray interpretations
very complicated and limit the applicability of Method A. However,
Method B is still of the simplicity in mathematics and is suitable
for multilayered structures.

The NFFF boundary in the multilayered structures is mainly affected
by the lateral waves as well. The attenuation behaviors of these
waves are similar to the ones shown in Fig. 4, so that Eq. (9) is
correct in the order of magnitude. The value of the dimensional
parameter $D$ depends on whether the VED is in the top layer or in
the intermediated region. When the VED is in the top layer, $D$
equals the distance between the VED and the lowest interface at
$z=d_{Q-1}$; when the VED is in one of the intermediate layer, $D$
is the distance between the highest interface at $z=d_{1}$ and
lowest interface at $z=d_{Q-1}$, which is the total thickness of the
intermediate region. The multi-reflection processes do not affect
the $D$ value for the following two reasons. Firstly, the
multi-reflections have a quick convergence. All the results of
Method A shown in the figures converge when $m$ is up to $2\sim4$,
which does not change the order of $D$. Secondly, as can be seen
from Fig. 4, the lateral waves have a nearly $r^{-1}$ decay when $D$
is small, and have a more rapid decay when $D$ is relatively larger.
Therefore the influence from the multi-reflection may decay to a
negligible one within the distance given by Eq. (9). Moreover, it is
this decay characteristic of the lateral waves that makes the NFFF
boundary almost independent of the wavelength.

For other field components of the dipole radiations and the cases of
different dipole orientations, we have numerically verified the
correctness of Eq. (9). Besides, all these cases depend on the
evaluations of the scattering coefficients related SIs, but the
field expressions are slightly different, which results in different
far-field patterns over the observation angles. However, in the
asymptotic analysis, they have the same SPs and similar asymptotic
expressions and branch cut contributions. Thus, Eq. (9) is
applicable.

In the beginning of Sec. 2, we have assumed that all of the layers
were lossless. Now let us discuss what about the case where the
intermediate region is not lossless. In such a situation, two cases
are distinguished to discuss the NFFF boundary: the loss is large or
small. When the loss is large, as in undersea communications where
short waves may not reach the sea bottom \cite{r22,r23,r24}, the
air-ocean-earth trilayer model simplifies to an air-ocean bilayer
one. From the discussions in Sec. 3, a distance of $10\lambda_{0}$
is enough to differentiate the near field and far field in air. When
the loss is relatively small, the amplitude of light decreases as it
arrives the bottom interface, which accordingly leads to a decrease
of the differences between the asymptotic results and the accurate
ones. Generally speaking, losses make the NFFF boundary reduced: its
value will be less than that given by Eq. (9).

Our conclusion can extend to bulk sources. On the one hand, within
the scope of volume integral method \cite{r08}, a bulk source can be
considered as a superposition of dipoles. The far-field radiations
of each dipole have been fully discussed above. On the other hand,
within the scope of NFFF transformation \cite{r07,r27,r28}, the near
fields of an arbitrary source can be converted to equivalence
surface dipoles on a virtual closed surface. Therefore the
dimensional parameter $D$ should anchor to the brightest spots in
the near field. Of course, our conclusions can be considered as an
applicable scope of the NFFF transformation.


\section{6. Concluding remarks}

In this paper, we have investigated the far-field asymptotic
behaviors of dipole radiations in stratified backgrounds and obtain
a universal empirical expression of NFFF boundary, Eq. (9). The
asymptotic results are compared with the accurate numerical ones in
various configurations to make sure the universality of Eq. (9). The
NFFF boundary is mainly affected by the lateral waves, which
correspond to branch point contributions in the SIs and are of a
higher-order attenuation. In Eq. (9), the parameter $D$ plays a key
role, and it is much larger than the operating wavelength. As a
result, the NFFF boundary in a stratified background is totally
different from that in vacuum. To be more specific, Eq. (9)
describes the boundary between the intermediate and far fields.
However, since the intermediate field is vaguely defined in optics,
we still call it as the NFFF boundary. In the case that the
observation point is in a region where its RI is the lowest in the
whole structure (usually air), the lateral wave decay rapidly, and
the NFFF boundary is about ten wavelengths.

It is believed that our conclusions are very helpful in
understanding and applying the far-field approximation. In
electromagnetic simulations, such as the finite-difference
time-domain method and the finite element method, the far-field
results are obtained by a NFFF transformation where the far field
approximation, or the LOA, is employed. The NFFF boundary presented
here reveals the applicability of the NFFF transformation,
especially in the forbidden region. Moreover, we compare the
different treatments for SPs in the asymptotic method and improve
the accuracy according to the ray theory (Method A). The equivalence
between the results of the asymptotic method and reciprocal theorem
is demonstrated.

\section*{Acknowledgement}

This work was supported by the China Postdoctoral Science Foundation
(Grant No. 2013M542222) and the National Natural Science Foundation
of China (Grant Nos. 11334015 and 61275028).



\begin{references}
\bibitem{r38} {Since the power series could not properly retrieve the lateral wave in the vicinity of the interface, the fitting range is $4000\sim10000\lambda_{0}$  which is sufficiently wide for accurate fitting.}
\end{references}
\end{document}